\documentclass[aip,jmp,reprint,showpacs,preprintnumbers,onecolumn]{revtex4-1}

\usepackage{amsfonts,amsthm,amsmath,amssymb} 
\usepackage{bm,bbm}
\usepackage{graphicx}
\usepackage[T1]{fontenc}
\usepackage{esint}
\usepackage{color}
\usepackage{braket}
\usepackage{hyperref}

\newcommand{\be}{\begin{equation}}
\newcommand{\ee}{\end{equation}}
\newcommand{\ba}{\begin{eqnarray}}
\newcommand{\ea}{\end{eqnarray}}
\def\bs{\begin{subequations}}
\def\es{\end{subequations}}
\def\a{\alpha}
\def\b{\beta}

\def\e{\epsilon}

\def\om{\omega}

\def\s{\sigma}
\def\vr{\varrho}
\def\vp{\varphi}

\def\cD{{\cal D}}

\def\cK{{\cal K}}

\def\cV{{\cal V}}
\def\bE{\mathbbm{e}}
\def\ds{d_{\rm S}}
\def\dh{d_{\rm H}}

\def\lh{l_{\rm H}}
\def\p{\partial}
\def\bp{\bar{\partial}}

\newcommand{\Eq}[1]{(\ref{#1})}
\def\com{\color{magenta}}
\def\cob{\color{blue}}
\newcommand{\arX}[1]{\href{http://arxiv.org/abs/#1}{{\ttfamily\com arXiv:#1}}}
\newcommand{\doin}[5]{\href{http://dx.doi.org/#1}{\cob {#2} \textbf{#3}, #4 (#5)}}
\newcommand{\ndoin}[5]{\href{#1}{\cob {#2} \textbf{#3}, #4 (#5)}}
\newcommand{\boxd}[1]{\boxed{\phantom{\Biggl(}#1\phantom{\Biggl)}}}
\newcommand{\tia}[1]{}

\def\rme{e}
\def\rmd{d}
\def\rmi{i}

\begin{document}

\title{Quantum mechanics in fractional and other anomalous spacetimes}

\author{Gianluca Calcagni}
\email{calcagni@iem.cfmac.csic.es}
\affiliation{Max Planck Institute for Gravitational Physics (Albert Einstein Institute)\\
Am M\"uhlenberg 1, D-14476 Golm, Germany}
\affiliation{Instituto de Estructura de la Materia, CSIC, Serrano 121, 28006 Madrid, Spain}
\author{Giuseppe Nardelli}
\email{nardelli@dmf.unicatt.it}
\affiliation{Dipartimento di Matematica e Fisica, Universit\`a Cattolica, via Musei 41,\\ 25121 Brescia, Italy}
\affiliation{INFN Gruppo Collegato di Trento, Universit\`a di Trento, 38100 Povo (Trento), Italy}
\author{Marco Scalisi}
\email{m.scalisi@rug.nl}
\affiliation{Max Planck Institute for Gravitational Physics (Albert Einstein Institute)\\
Am M\"uhlenberg 1, D-14476 Golm, Germany}
\affiliation{Centre for Theoretical Physics, University of Groningen, Nijenborgh 4,\\ 9747 AG Groningen, The Netherlands}

\begin{abstract}
We formulate quantum mechanics in spacetimes with real-order fractional geometry and more general factorizable measures. In spacetimes where coordinates and momenta span the whole real line, Heisenberg's principle is proven and the wave-functions minimizing the uncertainty are found. In spite of the fact that ordinary time and spatial translations are broken and the dynamics is not unitary, the theory is in one-to-one correspondence with a unitary one, thus allowing us to employ standard tools of analysis. These features are illustrated in the examples of the free particle and the harmonic oscillator. While fractional (and the more general anomalous-spacetime) free models are formally indistinguishable from ordinary ones at the classical level, at the quantum level they differ both in the Hilbert space and for a topological term fixing the classical action in the path integral formulation. Thus, all non-unitarity in fractional quantum dynamics is encoded in a contribution depending only on the initial and final state.
\end{abstract}

\date{July 18, 2012}


\preprint{\arX{1207.4473}\hspace{14cm} AEI-2012-044}
\begin{center}
\preprint{\doin{10.1063/1.4757647}{JOURNAL OF MATHEMATICAL PHYSICS}{53}{102110}{2012}\hspace{5cm}}
\end{center}


\maketitle


\section{Introduction}

In the attempt to understand the relation between the ultraviolet (UV) finiteness of different theories of quantum gravity and their geometric properties, a framework has been formulated where the geometry of spacetime displays most of the characteristics of multi-fractals, with an anomalous and scale-dependent dimensionality. A traditional perturbative field theory lives in a background geometry described by fractional calculus.\cite{fra4,frc1,frc2,ACOS,frc3,fra6,frc4} Renormalizability is achieved, at least at the level of power counting, thanks to the change of correlation functions with the scale, from a four-dimensional infrared down to small scales where spacetime is effectively two-dimensional. This is a general property of models in spacetimes with Lebesgue--Stieltjes measures,\cite{fra1,fra2,fra3} which are particular realizations of dimensional flow in quantum gravity.\cite{tH93,Car09,Car10} Contrary to general Lebesgue--Stieltjes measures, the coordinate dependence of fractional measures is factorized,\cite{frc1,frc2} which crucially allows one to define momentum space and a ``Fourier'' transform.\cite{frc3} Also, fractional measures make it possible to realize multi-scale geometries in a quantitative, controlled way.\cite{fra6,frc4}

Depending on its interpretation and applications, multi-fractional theory constitutes either a fundamental or effective framework. In the second case, it aims to describe the geometry of other models through the tools of multi-fractal geometry, transport and stochastic theory. These tools are generic enough to allow for a ``portable'' description of the physics across the models. For instance, the effective cyclic-invariant measures in classical coordinates of $\kappa$-Minkowski and other non-commutative spacetimes with non-linear algebras are of fractional type and link non-commutative and fractal geometries.\cite{ACOS} Also, the fractal properties of the renormalization group flow in asymptotic safety \cite{RSnax} are reproduced via fractional multi-scale diffusion equations.

Taken as a fundamental description of spacetime, on the other hand, fractional geometry is promising but a number of elementary questions still need to be addressed. A very basic one is, in fact, whether quantum mechanics exists in such geometries and, if it does, how it is modified with respect to the ordinary case. The presence of a non-trivial measure breaking all Poincar\'e symmetries and with a singularity at the coordinate axes may in fact carry unforeseen consequences for quantization (first as well as second) and the structure of Hilbert spaces and operators. 

The purpose of this paper is to answer the above question in the affirmative and show that quantum mechanics on fractional spacetimes is well defined. Thanks to a simple field redefinition, the case of a free non-relativistic particle can be formally recast as the one in standard geometry, the only difference at the canonical level being in the profiles of the wave-functions. As a consequence, even the free theory is non-trivial. It is also very easy to work out the quantum fractional harmonic oscillator in all its details. For interacting systems (non-linear equations of motion), the underlying genuine fractal/anomalous nature of space and time is further stressed by the fact that the measure weight cannot be reabsorbed completely into the position variables. 

Although ordinary time and space translations are lost, there survives a modified notion of them which highlights the existence of a very convenient mapping of the dynamics into a unitary one. Even if the dynamics is non-unitary due to the typical dissipative nature of fractional models, we shall still be able to introduce self-adjoint operators, manipulate probabilities and define scalar products and temporal evolution in almost the same fashion as standard quantum mechanics. This is possible for the following reason. In conventional dissipative systems, signals decay in time and space and states cannot evolve unitarily. This means that there do not exist observables associated with a self-adjoint operator $\hat{\bar P}=-\rmi\hbar \p_x$ generating spatial translations and with a Hamiltonian $\hat{\bar H}$ generating time translations. The problem is circumvented in the fractional scenario because both space and time are modified in such a way that one can construct momentum and Hamiltonian operators $\hat P\neq \hat{\bar P}$ and $\hat H\neq \hat{\bar H}$ which are not self-adjoint with respect to the standard scalar product, but that are self-adjoint with respect to the natural scalar product on fractional spaces. Thus, one obtains a notion of conserved scalar product and eigenfunctions of stationary states do not decay. To achieve this, it is sufficient to notice a one-to-one correspondence between operators $\hat{\bar A}$ of the ordinary theory and operators $\hat A$ of the one with factorizable (in particular, fractional) measure. In position space,
\be
\hat{\bar A}\quad \longleftrightarrow\quad \hat A:= \frac{1}{\sqrt{v_i(x)}}\,\hat{\bar A} \sqrt{v_i(x)}\,,
\ee
where $v_i(x)$ is the spatial part of the non-trivial measure weight. For the evolution operator $\hat U$ from time $t'$ to $t$, there is also an extra time dependence in the form $\sqrt{v_0(t')/v_0(t)}$, where $v_0$ is the time-dependent part of the measure weight. Although the theory is non-unitary, the $S$-matrix limit of the fractional propagator is unitary due to trivialization of the term $\sqrt{v_0(t')/v_0(t)}$ in the double limit $t'\to-\infty$, $t\to+\infty$, provided $v_0$ depends on $|t|$. Furthermore, it is shown that the classical action $S_{\rm cl}$ evaluated on classical trajectories is the same in ordinary and fractional theory. Nevertheless, $S_{\rm cl}$ enters the path integral formulation in a rather interesting way, i.e., augmented by a topological term distinguishing the ordinary from the fractional path integral; this term is, again, nothing but a rewriting of the ratio $\sqrt{v_0(t')/v_0(t)}$. All these features will be discussed in due course.

After briefly recalling the main properties of fractional geometries and extending the discussion to even more general measures (Sec.\ \ref{bafa}), in Sec.\ \ref{quame} 
 we motivate bilateral fractional spaces (introduced in Ref.\ \onlinecite{frc3}) by the interpretational difficulties arising in quantum mechanics if both coordinates and momenta are positive definite. We define a pair of self-adjoint operators (position and momentum) for quantum mechanics and prove Heisenberg's uncertainty principle. The latter, which is formally the same as in the integer case, is minimized by a weighted version of Gaussian functions. The Schr\"odinger equation and the operator of time evolution are then constructed. The classical and quantum free particle and the harmonic oscillator are studied in Sec.\ \ref{exe}, while the path integral formulation on fractional spaces is developed in Sec.\ \ref{path}. Conclusions are in Sec. \ref{conc}.

Before starting, we comment on other works falling under the unrelated but almost homonymous topic of ``fractional quantum mechanics.'' It is known that the fine-scale structure of a quantum mechanical particle path is very irregular and described by a nowhere-differentiable curve akin to a Brownian path.\cite{FeH,AbW} (This prompted an early attempt to formulate a model of quantum mechanics in fractal spacetime,\cite{Ord83} not based on fractional calculus.) As normal diffusion associated with a Wiener process is only one among many possibilities realized by Nature, and anomalous diffusion described by fractional equations is the rule rather than the exception (see Refs.\ \onlinecite{frc4,MeK,Zas3,MeK2} and references therein), it is spontaneous to ask what type of quantum mechanics would emerge from ``anomalous'' paths. Anomalous transport takes place in chaotic quantum systems, of which fractional dynamics is a model in certain regimes.\cite{KBD1,KBD2} Breaking of time reversal and dissipation typical of classical fractional mechanics translate to a non-conservation of probability in the quantum realm. These facts prompted curiosity in formulating first-quantization dynamics with fractional derivatives. The Schr\"odinger equation was generalized to the case where the Hamiltonian $\hat H$ features a spatial Riesz Laplacian \cite{Wes00,Las1,Las2,Las3,Her05,GX,Iom07,DX1,Las4} (stemming from a path integral over L\'evy paths); long-range interactions in quantum lattices can be modeled by this type of dynamics.\cite{LaZ} Other modifications of the Schr\"odinger equation involve a non-Markovian time evolution of the form $[(\rmi\hbar)^\b\p_t^\b-\hat H]\psi=0$, where $\p_t^\b$ is the fractional derivative of order $0<\b\leq 1$ \cite{Nab04,Bha07,OMA} (in particular, when $\b=1/2$ fractional dynamics can be described by a comb model \cite{BaI,Iom09}), or both time and space fractional derivatives.\cite{DX2,WX} See Ref.\ \onlinecite{Tar08} for another type of heuristic application of fractional operators to quantum mechanics.

Our case is quite different with respect to both motivation (as stated above) and techniques. The framework we present does not describe anomalous quantum systems in an ordinary spacetime but ordinary quantum systems in an intrinsically anomalous (in particular, fractal) spacetime. At a loose phenomenological level, this interpretation might also hold for some of the cited works, but our technical implementation is unique to fractional spacetimes as declared in Refs.\ \onlinecite{fra4,frc1,frc2,frc3}. Morevover, our Laplacian is not a fractional operator but a plain second-order operator with fractional weights, chosen from the three requirements of (i) reduction to two dimensions in the ultraviolet of a quantum field theory,\cite{fra4,frc2} (ii) of being a quadratic form and (iii) of self-adjointness.\cite{frc3} 


\subsection{Basic facts about fractional spacetimes}\label{bafa}

Fractional spacetimes are a continuum with anomalous properties realizing fractal as well as non-fractal types of geometry. They have been introduced in Refs.\ \onlinecite{fra4,frc1} having a field theory in mind, so they are formulated as a redefinition of the Lebesgue measure of time and space. Here we limit to a minimal set of definitions sufficient to the scope of the present paper. In $D$ topological dimensions, $\mu=0,1,\dots,D-1$ (where $x^0=t$ is time), the measure $\rmd^Dx=\rmd t\,\rmd^{D-1}{\bf x}$ is replaced by
\bs\label{frme}\ba
&& \rmd\vr_\a(x):=\rmd^Dx\,v_\a(x):=\rmd t\, v_{\a_0}(t)\prod_{i=1}^{D-1}\rmd x^i v_{\a_i}(x_i),\\
&& v_{\a_\mu}(x^\mu):=\frac{|x^\mu|^{\a_\mu-1}}{\Gamma(\a_\mu)}\,,
\ea\es
where $0<\a_\mu\leq 1$ are $D$ constants dubbed fractional charges and $\Gamma$ is the gamma function. There exists also a more fundamental definition of $\vr_\a$ based on complex fractional charges, which leads to a discrete spacetime texture \cite{fra4,frc2} and is closely related to $\kappa$-Minkowski non-commutative geometry,\cite{ACOS} but we shall not consider it here. The real-order measure \Eq{frme} is said to be isotropic if all $\a_\mu=\a$ are equal. Without loss of generality, we consider a spatially isotropic configuration where all the spatial charges are equal, $\a_i=\a$, but the time-direction charge $\a_0$ may differ from them.

We distinguish between unilateral measures where $x^\mu\geq 0$ and bilateral measures where the support of $v_\a$ is the whole real line (minus the axes $x^\mu=0$). In the first case, the absolute value in \Eq{frme} is pleonastic and the world is limited to the first coordinate orthant. 

Clearly, the Poincar\'e group is completely broken. In general, one could define a non-factorizable Lebesgue--Stieltjes measure weight $v(t,{\bf x})$ preserving at least some of the symmetries, as in a previous attempt to realize anomalous spacetimes with a rotation-invariant measure.\cite{fra1,fra2,fra3} However, factorizability of the coordinate dependence is crucial for the definition of the momentum transform in $D$ dimensions, which is \cite{frc3}
\bs\label{fumt}\ba
\tilde f(k) &:=& \int_0^{+\infty}\rmd\vr_\a(x)\, f(x)\,c_{\a,\a'}(k,x)\,,\\
f(x) &=& \int_0^{+\infty}\rmd\vr_{\a'}(k)\,\tilde f(k)\,c_{\a,\a'}(k,x)\,,\\
c_{\a,\a'}(k,x)&:=&\left(\frac{2}{\pi}\right)^{\frac{D}{2}}\frac{1}{\sqrt{v_{\a'}(k)v_\a(x)}}\prod_\mu\cos(k^\mu x^\mu)\,,\nonumber\\
\ea\es
for the unilateral world, and
\bs\label{fbmt}\ba
\tilde f(k) &:=& \int_{-\infty}^{+\infty}\rmd\vr_\a(x)\,f(x)\,\bE_{\a,\a'}^*(k,x)\,,\label{fbit}\\
f(x) &=& \int_{-\infty}^{+\infty}\rmd\vr_{\a'}(k)\,\tilde f(k)\,\bE_{\a,\a'}(k,x)\,,\label{fstaue}\\
\bE_{\a,\a'}(k,x)&:=&\frac{1}{\sqrt{v_{\a'}(k)v_\a(x)}}\,\frac{\rme^{\rmi k\cdot x}}{(2\pi)^{\frac{D}{2}}}\,,\label{ea12}
\ea\es
for the bilateral world. This is only one choice among an infinite class of transforms and we have allowed for a momentum-space measure different from position space. Both $c_{\a,\a'}$ and $\bE_{\a,\a'}$ are eigenfunctions of the fractional Laplacian (in Euclidean signature)
\be
\cK_\a:=\sum_\mu\cD_\mu^2\,,\quad\qquad \cD_\mu := \frac{1}{\sqrt{v_\a(x)}}\, \p_\mu \left[\sqrt{v_\a(x)}\,\,\cdot\,\right]\,,\label{DD}
\ee
with eigenvalue $-|k|^2$. For simplicity, we shall concentrate on this second-order operator, although operators of fractional order are possible.\cite{frc4}

Giving up factorizability in a generic weight $v(t,{\bf x})$ quickly leads to a technically untractable model. Here, we shall further stress this point in several occasions with novel arguments. Moreover, contrary to the non-factorizable case fractional measures favour control over all details, including the rigorous determination of the Hausdorff, spectral and walk dimension of spacetime \cite{frc1,frc4} and the formulation of a multi-scale geometry in direct contact with the definitions in complex systems.\cite{frc2,frc4} 

Regarding the dimensions, the volume of a $D$-ball with radius $R$ scales as $\cV^{(D)}\sim R^{\dh}$, where
\be
\dh=\sum_\mu\a_\mu
\ee
is the Hausdorff dimension.\cite{frc1} For our particular choice of spatial isotropy, $\dh=\a_0+(D-1)\a$. On the other hand, the spectral dimension $\ds$ depends on the type of Laplacian and diffusion associated with spacetime. In the simplest case of normal diffusion, $\ds=\dh$ and the walk dimension is exactly equal to 2.\cite{frc4}

The physical picture envisages a spacetime which is effectively two-dimensional at small scales, so that field theories therein have improved UV behaviour,\cite{frc2} while at large scales one must recover a spacetime with Poincar\'e symmetries restored and the ordinary integer dimensionality $\dh=\ds=D$. To realize this picture of ``dimensional flow,'' one needs to upgrade to a multi-scale geometry and sum over all possible $\a$'s, so that the total measure weight is of the form $\sum_n g_n v_{\a_n}$ for some dimensionful couplings $g_n$.\cite{frc2,frc4} This definition is based upon the standard one of multi-fractal geometry. For instance, in an isotropic setting the minimum set giving rise to multi-scale behaviour is $\a\in\{\a_*,1\}$, where $\a_*=2/D$ is the UV fractional charge. A direct manipulation of multi-fractional measures can become rather complicated, also because they may be no longer factorizable and it is not clear whether it is possible to define a momentum transform across all the scales.\cite{frc3} Therefore, it is often simpler to work at different ``snapshots'' of the geometry by considering a measure with fixed, constant charges $\a_\mu$.

Thanks to the resiliency of the formalism, we can generalize the discussion to factorizable measures which are not fractional, by leaving the \emph{Ansatz}
\be\label{genf}
\rmd t\rmd^{D-1}{\bf x}\quad \longrightarrow \quad \rmd \vr(x)=\rmd t\, v_0(t)\,\prod_{i=1}^{D-1}\rmd x^i\,v_i(x^i)
\ee
completely unspecified. At one's own convenience, one can replace fractional measures with
this general form and \emph{vice versa}, 
\be
v_{\a_0}\leftrightarrow v_0\,,\qquad v_{\a_i}\leftrightarrow v_i\,.
\ee
As an added bonus, we may even regard $v_\mu(x)=\sum_n g_n v_{\a_{\mu,n}}(x)$ as an element of a factorizable \emph{multi}-fractional measure. For simplicity we shall continue to call the measures \Eq{genf} fractional, but the reader should keep in mind this important generalization.


\section{Quantum mechanics}\label{quame}

Once defined a unitary momentum transform $F_v$ on a suitable $L^2$ space,\cite{frc3} it is natural to ask whether it is possible to formulate quantum mechanics on fractional spacetimes and, more generally, on spacetimes with factorizable measure. 
We shall:
\begin{enumerate}
\item[(i)] define a pair of self-adjoint  operators $\hat Q$ and $\hat P$ obeying the Heisenberg algebra;
\item[(ii)] show that $\hat Q$ and $\hat P$ are related to each other by the momentum transform $F_v$, which is an isomorphism between two equivalent Hilbert spaces (spanned by $x$ and $k$);
\item[(iii)] check whether Heisenberg's uncertainty principle holds;
\item[(iv)] find a complete basis of the function space as eigenstates of a self-adjoint operator;
\item[(v)] discuss time evolution, Schr\"odinger equation, and the Green function.
\end{enumerate}
Applications of the formalism to the free particle and the harmonic oscillator will be given in Sec.\ \ref{exe}.

In unilateral worlds, the momentum transform is real, hence wave-functions are real, too, and there is no self-adjoint first-order derivative operator. (Every integration by parts entails a minus factor, so that, at best, a first-order derivative could be an anti-hermitian operator. The first available self-adjoint operator would then be second-order, which would have the chance to be self-adjoint only because two integrations by parts restore the correct sign.) Moreover, a world with only positive-definite momenta would be problematic: particles could only run away from the origin, without stopping or coming back, and one would be unable to construct bound states and to confine particles in a natural way. This is quite different from the ordinary situation of a particle with an infinite potential wall at $x\leq 0$, because in that case momentum space still would be the whole real line. In quantum field theory, one would face similar paradoxes: Feynman diagrams would lose their meaning, as all external legs in a diagram would carry only incoming momenta, the concept of anti-particle would be missing, and so on. Looking back at the derivation of the unilateral transform,\cite{frc3} it is more sensible to interpret both $x$ and $k$ as, actually, the absolute values $|x|$ and $|k|$ of a bilateral world. Therefore, we regard the bilateral world as the correct representation of fractional geometries where one can attempt to do meaningful physics. (Since $k^2$ is positive definite, the derivation of the Green function of a scalar field in Ref.\ \onlinecite{frc2}, worked out in the unilateral case, is unaffected. The only difference is that, in a bilateral world, the spectrum is twice that of the unilateral one.)


\subsection{Hilbert space}

From here on we specialize to one dimension (and still call $v_i$ the spatial measure weight). Let $L_\vr^2(\mathbb{R})$ be the Hilbert space of functions $\psi$ which are square integrable with respect to the non-trivial measure. This space is equipped with the inner product
\ba
(\psi_1,\,\psi_2)&:=&\int_{-\infty}^{+\infty}\rmd\vr(x)\, \psi_1^*(x)\,\psi_2(x)\,,\qquad \psi_{1,2}\in L^2_\vr\nonumber\\
&=&\int_{-\infty}^{+\infty}\rmd x\, v_i(x)\, \psi_1^*(x)\,\psi_2(x)\,.\label{scpr}
\ea
 We shall often employ the bra-ket notation $\braket{\psi_1|\psi_2}$ and reserve the symbol $\braket{\psi_1|\psi_2}_0:=\int\rmd x\,\psi_1^*(x)\,\psi_2(x)$ for the scalar product with respect to the ordinary Lebesgue measure.

The existence and finiteness of the norm $\|\psi\|^2:=(\psi,\,\psi)$ requires that
\bs\label{psps}\ba
\psi(x) &\ \stackrel{x\to\infty}{\sim}\ & x^{-\nu}\,,\qquad {\rm Re}(\nu)>\frac{\a}{2}\,,\\
\psi(x) &\ \stackrel{x\to 0}{\sim}\ & x^{-\rho}\,,\qquad {\rm Re}(\rho)<\frac{\a}{2}\,.
\ea\es
Next, we define two self-adjoint operators on this space. For  the position operator, we shall choose
\be
\hat Q:=x\,,
\ee
with domain  $D_Q=\{ \psi \in L_\vr^2 ~:~ x\psi \in L_\vr^2\}$. Looking at Eq.~\Eq{DD}, it is natural to define as momentum operator
\be
\label{hatp}
\hat P:=-\rmi\hbar\cD_x=-\rmi\hbar \frac{1}{\sqrt{v_i(x)}}\, \p_x \left[\sqrt{v_i(x)}\,\,\cdot\,\right]\,.
\ee
We establish the domain $D_P$ by requiring $\hat P$ to be self-adjoint. First of all, it will be obviously  necessary that $\sqrt{v_i(x)}\,\psi(x) \in C^1(\mathbb{R})$, in order to perform derivatives. Notice that the existence of the derivative is required for the function $\sqrt{v_i}\,\psi$, but \emph{not} for $\psi$. Secondly, we should demand that the image of $\hat P$ is in $L^2_\vr$, $\hat P \psi(x) \in L_\vr^2$. Finally, we need to check the absence of boundary terms when performing integrations by parts. Let us choose for now $\psi_a\in L_\vr^2$, $a=1,2$, such that $\sqrt{v_i}\, \psi_a \in C^1(\mathbb{R})$. Then,
\ba
(\psi_1,\, \hat P\psi_2) &=& -\rmi\hbar\int_{-\infty}^{+\infty}\rmd\vr(x)\, \psi_1^*(x)\,\cD_x\psi_2(x)\nonumber\\
                         &=& -\rmi\hbar\int_{-\infty}^{+\infty}\rmd x \sqrt{v_i(x)}\,\psi_1^*(x)\,\p_x[\sqrt{v_i(x)}\,\psi_2(x)]\, \nonumber\\
                         &=& -\rmi\hbar\left[v_i(x)\psi_1^*(x)\psi_2(x)\right]\Big|_{-\infty}^{+\infty}+\rmi\hbar\int_{-\infty}^{+\infty}\rmd\vr(x)\, \cD_x\psi_1^*(x)\,\psi_2(x)\nonumber\\
                         &=& \int_{-\infty}^{+\infty}\rmd\vr(x)\, [-\rmi\hbar\cD_x\psi_1(x)]^*\,\psi_2(x)\nonumber\\
                         &=& (\hat P\psi_1,\, \psi_2)\,.
\ea
The boundary term in the third equality vanishes at infinity because $\psi_i\in L^2_\vr$. Note that,  due to the presence of  the (mild) singularity at $x=0$  in the fractional measure $v_i(x)=v_\a(x)$, the behaviour of $v_\a(x)\psi_1^*(x)\psi_2(x)$ has to be checked also at the origin: it could be either singular or discontinuous. However, since $\sqrt{v_\a}\, \psi_a \in C^1(\mathbb{R})$, then $v_\a(x)\psi_1^*(x)\psi_2(x)$ is such (as product of two $C^1(\mathbb{R})$ functions) and \emph{a fortiori} continuous in $\mathbb{R}$ and, in particular, at the origin. In general, the boundary term indeed vanishes, and with the choice of domain  $D_P=\{\psi\in L_\vr^2 ~:~ \sqrt{v_i(x)}\, \psi (x) \in C^1(\mathbb{R}),\ \hat P \psi (x) \in L_\vr^2 \}$ the operator $\hat P$ is self-adjoint.


\subsection{Position and momentum operators}

In the common domain $D_Q\cap  D_P$, the operators $\hat Q$ and $\hat P$ satisfy the Heisenberg algebra
\be\label{px}
[\hat Q,\hat P]=\rmi\hbar\,.
\ee
In a fractional space, the generalized eigenfunctions of the position operator $\hat{Q}$ are the fractional Dirac distribution $\delta_v$,
\be
\hat{Q}\delta_v(q,x)=q\,\delta_v(q,x)\,,
\ee
where \cite{frc3}
\be
\label{deltafrac}
\delta_v(x,x'):=\frac{\delta(x-x')}{\sqrt{v_i(x)v_i(x')}}
\ee
is consistent with the resolution of the identity provided by the transform \Eq{fbmt}. 


The normalized eigenfunctions of $\hat P$, with real eigenvalue $p=\hbar k\in \mathbb{R}$, are given by Eq.~\Eq{ea12}. We set $F_v$ to be an automorphism ($\a'=\a$ in fractional measures), which simplifies the discussion; a generalization is straightforward. Then,
\be
\label{aval1}
\hat P \bE_v(kx)= p\,\bE_v(kx)\,,\qquad \bE_v(kx)=\frac{1}{\sqrt{v_i(k)v_i(x)}}\,\frac{\rme^{\rmi kx}}{\sqrt{2\pi}}\,.
\ee
Since the functions $\bE_v (kx)$ are symmetric under the exchange $k\leftrightarrow x$, Eq.~\Eq{aval1} implies also
\be
\label{aval2}
-\rmi\frac{1}{\sqrt{v_i(k)}}\, \p_k \left[\sqrt{v_i(k)}\,\bE_v (kx) \right] = x\,\bE_v(kx)\,.
\ee
Equations \Eq{aval1} and \Eq{aval2} establish the isomorphism between the two equivalent Hilbert spaces $L_\vr^2(x)$ and $L_\vr^2(k)$ related by the (unitary) momentum transform \Eq{fbmt}, $F_v:L_\vr^2(x) \to L_\vr^2(k)$. In fact, once functions are decomposed according to \Eq{fbmt}, Eqs.\ \Eq{aval1} and \Eq{aval2} show that the operators $\hat P$ and  $\hat Q$ in $L_\vr^2(x)$ can be equivalently realized as multiplicative (by $k$) and derivative (with respect to the variable $k$) operators in $L_\vr^2(k)$: in $L^2_\vr(k)$ the two operators exchange their roles, and also their domains are  mapped one into the other. It should be noted that the request of having  self-adjoint differential operators severely restricts the form of the momentum operator in the fractional context and, in particular, it rules out naive fractional derivatives. If $\hat P$ were defined by a fractional derivative of some order $\b$, integration by parts would yield an infinite number of terms due a complicated Leibniz rule (Eq.\ (2.66) of Ref.\ \onlinecite{frc1}). Even in the ``trivial'' $\a=1$ case, one would still have a further problem: $(\psi_1,\,\p^\b \psi_2)=(\bp^\b\psi_1,\,\psi_2)$, where $\bp^\b$ is the right derivative operator. Mixed fractional operators of the form $a\p^\b+a^*\bp^\b$ exist which are self-adjoint,\cite{frc4} but we do not consider them here.

It is worth to stress that, for any given measure with a singularity at a point $\bar x$ (in this case, $\bar x=0$), $\hat P$ (hence $\hat H$ below) is uniquely defined because it is not translation invariant. This contrasts with ordinary quantum mechanics, where $\hat{\bar P}=-\rmi\hbar \p_x$ is translation-invariant.


\subsection{Heisenberg principle}

The Schwarz inequality
\be
|(\psi_1,\,\psi_2)|^2\leq \|\psi_1\|^2 \|\psi_2\|^2\,,
\ee
which holds in our vector space, is sufficient to prove Heisenberg's principle through the Robertson inequality. Define the expectation value of an hermitian operator $\hat A$ as $\langle \hat A\rangle:=(\psi,\hat A\psi)$, and the standard deviation about its mean value as $(\Delta A)^2:=\langle \hat A^2\rangle-\langle \hat A\rangle^2=\|(\hat A-\langle\hat A\rangle)\psi\|^2$. For any pair of hermitian operators $\hat A$ and $\hat B$, 
 apply the Schwarz inequality to the vectors $\psi_A=(\hat A-\langle\hat A\rangle)\psi$ and $\psi_B=(\hat B-\langle\hat B\rangle)\psi$, obtaining the Robertson inequality 
 $\|\psi_A\|^2\|\psi_B\|^2\ge [\langle [\hat A,\hat B]\rangle/(2\rmi)]^2$. When $\hat A=\hat P$ and $\hat B = \hat Q$, this becomes Heisenberg's uncertainty principle,
\be\label{heip}
(\Delta Q)^2(\Delta P)^2\geq \frac{\hbar^2}{4}\,,
\ee
just like in ordinary quantum mechanics. Any dependence from the non-trivial measure has been absorbed into the definitions of the position and momentum operators.


\subsection{Hamiltonian}

We can define a self-adjoint operator corresponding to the fractional quantum Hamiltonian of a point particle of mass $m$ in a real potential $V(x)$:
\ba 
\hat{H} &=&\frac{1}{2m}\hat{P}^2 + V(\hat{Q})\nonumber\\
&=&-\frac{\hbar^2}{2m}\cD^2_x + V(x)\nonumber\\
&=& - \frac{\hbar^2}{2m} \frac{1}{\sqrt{v_i(x)}}\p^2_x\left[\sqrt{v_i(x)}\,\cdot\,\right] + V(x)\,.
\ea
For every Hamiltonian of this form, one can always consider the following ordered relation:
\be\label{h0}
\hat{H}=\frac{1}{\sqrt{v_i(x)}}\,\hat{\bar H}\,\sqrt{v_i(x)}\,,
\ee
where $\hat{\bar H}=-[\hbar^2/(2m)]\p^2_x + V(x)$ is the corresponding Hamiltonian operator of the ordinary non-fractional quantum theory. By construction, $\hat H$ is self-adjoint.

The eigenvalue problem
\be
\hat{H}\psi_n(x)=E_n\psi_n(x)\,,
\ee
can be formulated as
\be
\hat{\bar H}\varphi_n(x)=E_n\varphi_n(x)\,,
\ee
where
\be\label{egnf}
\varphi_n(x):=\sqrt{v_i(x)}\psi_n(x)
\ee
are the eigenfunctions in the standard $L^2$ space. If the latter are orthonormal under the $L^2$ scalar product, so are the fractional eigenfunctions $\psi_n$ with respect to Eq.\ \Eq{scpr}, $\langle\psi_n|\psi_m\rangle=\delta_{nm}$.


\subsection{Time evolution and Schr\"odinger equation}\label{schr}

As usual in an ordinary quantum mechanical framework, we regard time as a parameter and consider the time evolution of the system. The use of a time with non-trivial fractional charge 
is strictly connected to some features of fractional spacetimes regarding violations of energy conservation.

In the Schr\"odinger picture, the natural generalization of the Schr\"odinger equation for a quantum mechanical state $\ket{\psi}$ is
\be\label{fse}
i\hbar\cD_t \ket{\psi(t)} =\hat{H}\ket{\psi(t)}\,,
\ee
where 
\be
\cD_t=\frac{1}{\sqrt{v_0(t)}}\p_t\left[\sqrt{v_0(t)}\,\cdot\right]\,.
\ee
The choice of $\cD_t$ in Eq.\ \Eq{fse} is dictated by naturalness (the classical system is defined via these derivatives, not $\p_t$) and guarantees that quantum mechanics retains the dissipative character of the classical theory (a Schr\"odinger equation defined with $\p_t$ instead of $\cD_t$ would lead to a unitary model).

As a side note, the imaginary-time version of the Schr\"odinger equation is sometimes regarded as the definition of the diffusion equation valid in a given space, where $\hat H$ is interpreted as the Laplacian $\cK$ and $t\to\s$ as an abstract diffusion time. The diffusion equation in fractional spacetimes typically features the ordinary first-order derivative $\p_\s$, or a fractional derivative in anomalous models, but never the diffusion operator $\cD_\s$ because $\s$ is an abstract parameter unrelated from the time fractional structure of the geometry.\cite{frc1,fra6,frc4} Thus, there is a contrast between quantum mechanics and field theory in fractional spacetimes, marked by a non-correspondence between Schr\"odinger and imaginary-time diffusion equation. Since the difference lies in geometric weights, this somewhat reminds one of the situation in general relativity, where the Laplacian is the Euclideanized Laplace--Beltrami operator and the diffusion equation is manifestly non-covariant.

Let us define, first, the quantum state
\be\label{phi}
\ket{\chi} := \sqrt{v_0(t)}\ket{\psi}\,,
\ee
and, then, the \textit{unitary} state:
\be
\ket{\varphi} := \sqrt{v_i(x)}\ket{\chi}=\sqrt{v_i(x)v_0(t)}\ket{\psi}\,.
\ee
Notice that $v_i(x)v_0(t)$ is the total fractional measure in $1+1$ dimensions and one could have obtained it from the start by defining both $\cD_t$ and $\hat H$ with a generic measure $v(t,x)$. Factorizability of the coordinate dependence in the measure plays a fundamental role in keeping time and space dependence separated and it is essential in order to define the eigenvalue problem consistently. Indeed, eigenfunctions of the Hamiltonian would depend on time in the case of a non-factorizable measure $v(t,x)$, as it is clear by looking at Eq.~\Eq{egnf}, and would lose their fundamental character as stationary states.

The state $\ket{\varphi}$ satisfies the usual (unitary) non-fractional Schr\"odinger equation
\be\label{nfse}
i\hbar\p_t \ket{\varphi(t)} =\hat{\bar H}\ket{\varphi(t)}\,,
\ee
where $\hat{\bar H}$ is defined in Eq.\ \Eq{h0}. Then, considering the complex conjugate of Eq.\ \Eq{nfse}, $-i\hbar\p_t \bra{\vp(t)}=\bra{\vp(t)}\hat{\bar H}^{\dag}$, and the requirement $\hat{\bar H}=\hat{\bar H}^{\dag}$, under the ordinary $L^2$ scalar product one gets $i\hbar\p_t \braket{\vp_1(t)|\vp_2(t)}_0= \braket{\vp_1(t)|\hat{\bar H}^{\dagger}|\vp_2(t)}_0- \braket{\vp_1(t)|\hat{\bar H}|\vp_2(t)}_0=0$, 
meaning that equal-time scalar products of the type  $\braket{\vp_1(t)|\vp_2(t)}_0$ are time independent, as expected.

By using the relation \Eq{phi}, we see that, in a fractional space, time evolution preserves $L_\vr^2$ products of the form
\ba 
\braket{\chi_1(t)|\chi_2(t)}&=&\braket{\psi_1(t)|v_0(t)|\psi_2(t)}\nonumber\\
&=&\braket{\psi_1(t')|v_0(t')|\psi_2(t')}\nonumber\\
&=&\braket{\chi_1(t')|\chi_2(t')}\,.\label{scpp}
\ea
This result lets us prefigure that time evolution in a fractional space and the associated operator are not unitary, unlike in ordinary quantum mechanics.

\subsubsection{Time evolution operator}

Let $\hat{\bar U}(t-t')=\exp[-(i/\hbar)\hat{\bar H}(t-t')]$ be the unitary operator associated with the time evolution of the state $\ket{\varphi}$. As it is well known, Eq.~\Eq{nfse} can be written as
\be\label{nftoe}
i\hbar\p_t \hat{\bar U}(t-t')\ket{\vp(t')} =\hat{\bar H}\hat{\bar U}(t-t')\ket{\vp(t')}\,,
\ee
and, simply by multiplying and dividing by appropriate measure factors, we get
\be
i\hbar\cD_t\left[ \frac{1}{\sqrt{v_i(x)v_0(t)}}\hat{\bar U}(t-t')\sqrt{v_i(x)v_0(t')}\ket{\psi(t')} \right]
=\hat{H}\frac{1}{\sqrt{v_i(x)v_0(t)}} \hat{\bar U}(t-t') \sqrt{v_i(x)v_0(t')}\ket{\psi(t')}\,,\nonumber
\ee
which is the analogue of Eq.~\Eq{nftoe} in a fractional space:
\be
i\hbar\cD_t \hat{U}(t,t')\ket{\psi(t')} =\hat{H}\hat{U}(t,t')\ket{\psi(t')}\,,
\ee
where $U(t,t')$ is the fractional time evolution operator defined as
\be
\hat{U}(t,t'):=\frac{1}{\sqrt{v_i(x)v_0(t)}} e^{-\frac{i}{\hbar}\hat{\bar H}(t-t')}\sqrt{v_i(x)v_0(t')}\,.\nonumber
\ee
Expanding the exponential operator and resumming, we get a form of $\hat{U}$ in which the fractional Hamiltonian $\hat{H}$ appears explicitly:
\begin{widetext}
\be\label{U}
\boxd{\hat{U}(t,t'):=\frac{1}{\sqrt{v_i(x)v_0(t)}} \hat{\bar U}(t-t') \sqrt{v_i(x)v_0(t')}= \sqrt{\frac{v_0(t')}{v_0(t)}}\,\rme^{-\frac{i}{\hbar}\hat{H}(t-t')}\,.}
\ee
\end{widetext}
The ordering of the factors in the last expression is irrelevant. Notice that we can rewrite this equation as
\be 
\hat{U}(t,t')=\rme^{-\frac{i}{\hbar}\hat{H}(t-t')-\frac12\int_{t'}^t\rmd\tau \ln v_0(\tau)}=\rme^{-\frac{i}{\hbar}\int_{t'}^t\rmd\tau[\hat{H}-\frac{i\hbar}2 \ln v_0(\tau)]}\,.
\ee
The simultaneous presence of non-trivial real and imaginary contributions to the evolution operator highlights the dissipative nature of the system. The $\ln v_0$ term will be further discussed in section \ref{topo}.

One can check that $\hat{U}$ still keeps some important features characterizing a time evolution operator:
\begin{itemize}
 \item The initial condition $\lim_{t\rightarrow t'} \ket{\psi(t)}=\ket{\psi(t')}$ is compatible with
      \be
      \lim_{t\rightarrow t'} \hat{U}(t,t')=\mathbbm{1}\,.
      \ee
 \item The composition law required by time evolution holds:
      \be \label{cl}
      \hat{U}(t,t')=\hat{U}(t,t'')\hat{U}(t'',t')\,,\qquad\ t'\leq t''\leq t\,.
      \ee
\end{itemize}

Nevertheless, the operator $\hat{U}$ shows some novelties characterizing time evolution in a fractional space:
\begin{itemize}
 \item Time-translation invariance is broken and $\hat{U}$ is not unitary:
       \be
      \hat{U}(t,t')\hat{U}^\dagger(t,t')=\hat{U}^\dagger(t,t')\hat{U}(t,t')=\frac{v_0(t')}{v_0(t)}\neq\mathbbm{1}\,,
       \ee
       so that
       \be
       \hat{U}^\dagger\neq\hat{U}^{-1}\,.
       \ee
       Indeed,
       \be
       \hat{U}^{-1}(t,t')=\rme^{\frac{i}{\hbar}\hat{H}(t-t')}\sqrt{\frac{v_0(t)}{v_0(t')}}= \hat{U}(t',t)
       \ee
       and
       \be
       \hat{U}^\dagger(t,t')= \hat{U}^{-1}(t,t')\frac{v_0(t')}{v_0(t)}\,.
       \ee
       When time-translating the system forwards and, then, moving it back to the initial instant, one gains a factor distinguishing the two states. This mechanism avoids the so-called ``grandfather paradox'' and introduces an arrow of time also at the microscopic level.

 \item $\hat{U}$ just implements what we found in Eq.\ \Eq{scpp}, that is, $L_\vr^2$ products such as $\braket{\psi_1|v_0|\psi_2}$ are preserved during time evolution:
       \ba
       \braket{\psi_1(t)|v_0(t)|\psi_2(t)}&=&\braket{\psi_1(t')|\hat{U}^\dagger(t,t') v_0(t)\hat{U}(t,t')|\psi_2(t')}\nonumber\\
       &=&\braket{\psi_1(t')|v_0(t')|\psi_2(t')}\,.
       \ea
       On the other hand, the scalar product $\braket{\psi_1|\psi_2}$ is \emph{not} preserved in time.
\end{itemize}
Thus, while the wave-function $\vp$ evolves unitarily the fractional wave-functions $\psi$ do not, 
\ba 
\psi(x,t)&=&\frac{1}{\sqrt{v_i(x)v_0(t)}}\int\rmd t'\,\hat{\bar U}(t-t')\vp(x,t')\nonumber\\
&=& \int\rmd t'\,\hat U(t,t')\psi(x,t') \,.
\ea





\subsection{Green function}\label{gf}

As we saw above, the evolution of a fractional quantum mechanical system is described by Eq.\ \Eq{fse}, $i\hbar\cD_t \psi(x,t) =\hat{H}\psi(x,t)$, with an initial condition $\psi(x,t)=\psi(x,t')$  at $t=t'$. From a mathematical point of view, we are dealing with a Cauchy problem. Therefore, also in the fractional case, the general solution can be obtained by means of the Green function $G(x,t;x',t')$ and, in particular, by solving the following system:
\be
\begin{cases}\label{fseg}
i\hbar\cD_t G(x,t;x',t') =\hat{H}G(x,t;x',t')\,,\\
G(x,t';x',t')= \delta_v(x,x')\,.
\end{cases}
\ee
The solution of Eq.~\Eq{fse} can be represented as
\be\label{psi}
\psi(x,t)= \int \rmd x'\,v_i(x')\, G(x,t;x',t')\, \psi(x',t')\,.
\ee

Following the same route outlined above, we rewrite Eq.~\Eq{fseg} as $i\hbar\p_t\bar G(x,t;x',t') = \hat{\bar H} \bar G(x,t;x',t')$, where $\bar G:= \sqrt{v_i(x)v_0(t)}G$ and $\hat{\bar H}$ are the unitary quantities of the ordinary quantum theory. The solution of the last equation is well known:
\be\nonumber
\bar G(x,t;x',t')=c \sum_n \varphi_n(x) \varphi^*_n(x') e^{-\frac{i}{\hbar}E_n (t-t')}\,,
\ee
where $c$ is independent of $x$ and $t$, and $\varphi_n$ are the eigenfunctions of the Hamiltonian $\hat{\bar H}$. The value $c=1$ corresponds to the initial condition $\bar G (x,t';x',t') = \delta (x-x')$. The form of $G$ is then 
\be
G(x,t;x',t')=\frac{c}{\sqrt{v_i(x)v_0(t)}} \sum_n \varphi_n(x) \varphi^*_n(x') e^{-\frac{i}{\hbar}E_n (t-t')}.
\ee
By imposing the initial condition of Eq.~\Eq{fseg}, we get the constant $c=\sqrt{v_0(t')/v_i(x')}$ and we can finally write the fractional Green function as
\be\label{gi}
\boxed{G(x,t;x',t')=\sqrt{\frac{v_0(t')}{v_0(t)}}\sum_n \psi_n(x) \psi^*_n(x') e^{-\frac{i}{\hbar}E_n (t-t')}\,,}
\ee
where we used the relation \Eq{egnf} and $\psi_n$ are the eigenfunctions of the fractional Hamiltonian $\hat{H}$. Again, factorizability of the measure is fundamental. Using the completeness relation of the eigenfunctions, it is straightforward to re-express Eq.\ \Eq{gi} via the time-evolution operator:
\be\label{xUx}
G(x,t;x',t')= \braket{x'|\hat{U}(t,t')|x}\,.
\ee
All the non-unitarity of the theory is encoded in the pre-factor $\sqrt{v_0(t')/v_0(t)}$. In Sec.\ \ref{topo} we shall provide a neat reinterpretation of this pre-factor as a topological term in the path integral formulation. For the time being, we notice that the $S$ matrix, defined as
\be 
\lim_{\substack{t'\to-\infty\\ t\to+\infty}}\braket{x'|\hat{U}(t,t')|x}\,,
\ee
\emph{is} unitary in fractional theories, provided the limits are performed symmetrically, so that $\sqrt{v_{\a_0}(t')/v_{\a_0}(t)}\to 1$. This is not true in general for theories with measures \Eq{genf}, even if $v_0$ is manifestly positive. An example is $v_0(t)=\exp(a t)$, where $a$ is a constant. A necessary and sufficient condition is $v_0$ to depend on $|t|$.


\section{Examples}\label{exe}

In this section, we shall develop the two basic quantum mechanical examples: the free particle and the harmonic oscillator.

Formally, a classical mechanics system is insensitive to the choice of spatial measure. Its fractional action is
\be
\label{actpart}
S=\int_{t_1}^{t_2} \rmd t\,v_0(t)\, L\,,
\ee
where only $v_0$ appears. In order to get the equation of motion, we apply  the variational principle by finding the stationary points of the action,
\be
\delta S= 0\,,
\ee
under variations $\delta q(t)$ preserving the extrema $\delta q(t_1)=\delta q(t_2)=0$, 
namely,
\be
\label{eulag1}
\int_{t_1}^{t_2}\rmd t\,v_0(t) \left(\frac{\p L}{\p\cD_tq}\delta\cD_tq + \frac{\p L}{\p q} \delta q\right) = 0\,.
\ee
Here we assumed that the Lagrangian depends only on $q$ and $\cD_tq$. Since
\ba
\delta\cD_tq&=& \cD_t(q+\delta q) -\cD_t(q)\nonumber\\
&=& \frac{1}{\sqrt{v_0(t)}} \p_t \left[\sqrt{v_0(t)} (q+\delta q) \right]-\frac{1}{\sqrt{v_0(t)}} \p_t \left[\sqrt{v_0(t)} q\right]\nonumber\\
&=& \frac{1}{\sqrt{v_0(t)}} \p_t \left[\sqrt{v_0(t)}\delta q\right]\nonumber\\
&=&\cD_t \delta q\,, 
\ea
integrating by parts Eq.\  \Eq{eulag1}  we have
\be
0=  - \int_{t_1}^{t_2} \rmd t\,v_0(t)\left[\cD_t \frac{\p L}{\p\cD_tq}-\frac{\p L}{\p q}\right]\delta q(t) + \left[ v_0(t)\frac{\p L}{\p\cD_tq} \delta q(t)  \right]_{t_1}^{t_2}\,.\label{eulag2}
\ee
The boundary term vanishes by virtue of $\delta q(t_1)=\delta q(t_2)=0$ and one is left with the fractional generalization of the Euler--Lagrange equation of motion
\be
\label{eulag3}
\cD_t\frac{\p L}{\p\cD_tq}- \frac{\p L}{\p q}=0\,.
\ee
Moving to Hamiltonian formalism, we define the conjugate momentum
\be
\label{p-clas}
p:=\frac{\p L}{\p\cD_tq},
\ee
which allows us to write the Hamiltonian as
\be
\label{H-clas}
H := p\cD_tq - L \, .
\ee
Upon postulating the canonical brackets $\{ q,p\}=1$ (the others vanish), fractional
Hamilton's equations are
\be
\cD_tq:=\{q,H\} \,,\qquad \cD_t p:=\{p,H\},
\ee
that are easily seen to be consistent with  \Eq{eulag3}. 


\subsection{Free particle}

\subsubsection{Classical particle}

The simplest example is that of a free particle. Its classical fractional Lagrangian is defined as
\be \label{freep}
L= \frac12 m (\cD_t q)^2\,,
\ee
leading to the Euler--Lagrange equation of motion 
\be
\cD_t^2 q = \frac{1}{\sqrt{v_0(t)}} \p^2_t \left[\sqrt{v_0(t)} q\right]=0\,.
\ee 
Notice that by introducing the variable (this setting is the classical analog of the quantum state $|\chi \rangle$ introduced in \Eq{phi})
\be\label{chiq}
\chi(t) = \sqrt{v_0(t)} q(t)\,,
\ee
the fractional equation of motion written in terms of $\chi$ would be indistinguishable from that of a free particle in a non-fractional space, i.e.,
\be\label{eomc} 
\ddot\chi =0\,.
\ee 
The classical fractional solutions are then easily written in terms of the non-fractional solutions $\chi(t) = \chi_0 + w t$, leading to
\be
q(t) = \frac{1}{\sqrt{v_0(t)}}\bigl(\chi_0+ w t\bigr)\,.
\ee
Since $\chi_0$ is an arbitrary constant, the first term of this equation highlights a symmetry of the Lagrangian \Eq{freep} in a time-dependent change of the position, $q(t)\to q(t)+\chi_0/\sqrt{v_0(t)}$, which reduces, for fractional measures, to an ordinary translation when $\a=1$.

The classical fractional action evaluated on the solutions of the equation of motion 
is identical to  that of the corresponding non-fractional system $\chi$,
\be
\label{sfree1}
S_{\rm cl}= \int_{t_1}^{t_2} \rmd t\,  v_0(t) \frac12 m (\cD_t q)^2 = \frac12 m w^2 (t_2-t_1)\,.
\ee
In spite of this fact, the  interpretation is different: $w$ is not the velocity (which is not constant) of the fractional variable $q(t)$.

The redefinition \Eq{chiq} and the corresponding trivialization of the problem obviously triggers the following remark. Since this (and also the harmonic oscillator below) is a free system, the measure weight can always be reabsorbed into the definition of the coordinates and the physics is classically equivalent to the usual one \emph{identically}. 

Therefore, there is seemingly no point in working out fractional rules (variational principle, time evolution, and so on) which turn out to be unnecessary, and fractional dynamics seems altogether trivial. Even for a non-linear system with a potential $V(q)$, one could simply consider an ordinary setting with a non-autonomous potential $f(t)V(\chi)$, for some function $f$. For instance, a power-law potential would yield $v_0(t)V(q)=v_0(t)q^n=v_0^{1-n/2}(t)\chi^n$. Are fractional spacetimes just trivial reformulations of non-autonomous systems?

At the classical level, the answer is mainly interpretational. A non-autonomous potential defined in an ordinary space can be considered by all means, but a justification for a particular choice should then be provided. So, while in ordinary spacetime a potential of the form $|t|^l\chi^n$ with $l>0$ would come out of the blue, its origin as an effective description of a system with fractional time and the subsequent relation $l=(n/2-1)(1-\a_0)$ between physical parameters are clear. In turn, fractional time and the specific form of the time-dependent parts of the system are grounded upon a rigorous implementation of fractal (in general, anomalous and multi-scale) geometry via fractional calculus,\cite{frc1,frc2,frc3,frc4} along the motivations outlined in the Introduction. The multi-scale anomalous/fractal picture and its tools cannot be forfeited without loosing logical coherence and the motivational drive.

However, this perspective is not particularly compelling in the case of free systems, where one can erase all measure dependence and work with $\chi$ coordinates. At the classical level, one could simply bypass the problem by adopting a Laplacian of fractional order \cite{frc4} and invoking again the interpretation according to fractal geometry. Unfortunately, this would not amount to an actual solution, because second-order Laplacians are desirable in a field-theory setting.\cite{frc2} Quantum mechanics comes to the rescue with a twofold simple answer. First, while the dynamics is formally reduced to the standard case due to the linearity of the Schr\"odinger equation, even free systems are not equivalent to non-fractional ones because the Hilbert space and the profile of wave-functions drastically change. This will be made clear in the example of the harmonic oscillator. Secondly, at the level of the path integral, the key difference between ordinary and fractional systems is a topological term, a total derivative immaterial in the classical theory.

\subsubsection{Quantum particle}

The quantum mechanical system can be easily obtained starting from the classical Hamiltonian that, according to \Eq{H-clas}, is
$H= p^2/(2m)$. A self-adjoint Hamiltonian operator $\hat H$ in $L^2_\vr$ is promptly obtained by replacing $p$ with the self-adjoint operator $\hat P$ as in \Eq{hatp}. Then the eigenfunctions of the Hamiltonian are  just \Eq{aval1}, with eigenvalue $\hbar^2 k^2/(2m)$. The spectrum is continuous, and the eigenfunctions $\psi_k(x)= \bE_v(kx)$ are normalized per unit volume.

The Green function of the system can be obtained from \Eq{gi} (with the obvious replacement $\sum_n \rightarrow \int \rmd k\, v_i(k)$ due to the continuous spectrum):
\begin{widetext} 
\ba
\label{gfree1}
G(x,t;x',t')&=& \sqrt{\frac{v_0(t')}{v_0(t)}}\int dk\,v_i(k)\, \bE_v(kx) \bE^*_v(kx')\,e^{-\frac{i \hbar}{2m} k^2 (t-t')}\nonumber \\
&=& \sqrt{\frac{v_0(t')}{v_0(t) v_i(x) v_i (x')}} \, \, \frac{1}{2\pi \hbar} \int dp\, e^{\frac{i}{\hbar} p (x-x')}\  e^{-\frac{i}{2m \hbar} p^2 (t-t')}\nonumber \\
&=& \sqrt{\frac{v_0(t')}{v_0(t) v_i(x) v_i (x')}} \, \,\sqrt{\frac{i m}{2 \pi \hbar (t-t') }} \  e^{\frac{i}{\hbar} m \frac{(x-x')^2}{2 (t-t')}},
\ea
\end{widetext} 
where, to perform the Gaussian integral, we used the regularization $m\to m+i \epsilon$ and took eventually the limit $\epsilon \to 0$. The argument of the exponential in the last equality of \Eq{gfree1} is just ($i/\hbar$)  times the action \Eq{sfree1}, that is $e^{(i/\hbar) S_{\rm cl}}$. Note, in fact, that the corresponding non-fractional  $\chi$ particle would have constant velocity $w= (x-x')/(t-t')$. Later we shall come back to the significance of the pre-factor $\sqrt{v_0(t')/v_0(t)}$.


\subsection{Harmonic oscillator}\label{ho}

As another application, we examine the harmonic oscillator. First, we build the classical fractional theory and obtain the equation of motion as well as the dynamical variables. Second, we move to the quantum theory, getting a complete orthonormal basis of $L^2_\vr$ by solving the associated eigenvalue problem.

\subsubsection{Classical  oscillator}

We define the fractional action  of the harmonic oscillator of mass $m$ and frequency $\om$ as
\ba 
S&=&\int_{t_1}^{t_2} dt\, v_0(t) \ L\nonumber\\
&=& \int_{t_1}^{t_2} dt\, v_0(t)\left[\frac{1}{2}m\left(\cD_tq\right)^2 - \frac{1}{2} m {\om^2} q^2 \right]\,,\label{lhar}
\ea
so that one gets the Euler--Lagrange equation
\be\label{eqho}
\cD^2_tq +{\om^2} q=0\,,
\ee
that is
\be
\ddot q+\frac{\dot v_0}{v_0}\dot q+\frac{1}{2}\left(\frac{\ddot v_0}{v_0}-\frac{1}{2}\frac{\dot v_0^2}{v_0^2}+{\om^2}\right)q=0\,.
\ee
In the fractional case $v_0(t)=|t|^{\a_0-1}/\Gamma(\a_0)$, we recognize an anti-friction and a potential term:
\be
\ddot q-\frac{1-\a_0}{t}\dot q+\left[\frac{(1-\a_0)(3-\a_0)}{4t^2}+{\om^2}\right]q=0\,.
\ee
For $\a_0=1$, we recover the standard non-fractional equation.

Moving to Hamiltonian formalism, we define the conjugate momentum
\be
p:=\frac{\p L}{\p\cD_tq}=m\cD_tq\,,
\ee
which allows us to write down the Hamiltonian as
\be
H := p\cD_tq - L = \frac{1}{2m}p^2+\frac{1}{2}m {\om^2}q^2\,.
\ee
 Fractional
Hamilton's equations  turn to be
\be
\cD_tq:=\{q,H\}=\frac{p}{m} \,,\qquad \cD_t p:=\{p,H\}= -m {\om^2} q\,,
\ee
where the definition of time evolution is consistent with the result \Eq{eqho}. Invariance of the dynamics under the exchange of $p$ and $q$ variables is evident in the case $\om=1/m$.

Like in the free-particle case, an equivalent way to get the same results is by defining a new coordinate variable $\chi:=\sqrt{v_0}q$ and a new Lagrangian $\bar L:=v_0L$. The dynamics is then the usual (non-fractional) one. The equation of motion one gets is indeed $\ddot\chi + \om^2\chi=0$, which is nothing but Eq.~\Eq{eqho}. The $\chi$ solutions are 
$\chi (t) = A \cos \om t + B \sin \om t$, from which the solution $q(t)$ trivially follows. By choosing boundary conditions $\chi_1 = \chi(t_1)$ and $\chi_2 = \chi(t_2)$, the $\chi(t)$ solution reads
\be
\label{chicl}
\chi(t) = \frac{1}{\sin[\om (t_2 - t_1) ]} \left\{\chi_2 \sin[\om (t - t_1) ]-\chi_1 \sin[\om (t - t_2)]\right\}\,,
\ee
and the classical action evaluated on the solution \Eq{chicl} reads
\be
S_{\rm cl}= \frac{m \om }{2 \sin[\om (t_2-t_1)]} \left\{(\chi_1^2 + \chi_2^2) \cos[\om (t_2 - t_1) ]-2\chi_1\chi_2\right\}\,.\label{sclh}
\ee
Just like in the free particle case, the classical action of the $\chi$ system \Eq{sclh} is 
 indistinguishable from the classical action $\bar S$ we would have obtained substituting the corresponding fractional solution $q(t) = \chi(t) /\sqrt{v_0 (t)}$ in the fractional action \Eq{lhar}.

\subsubsection{Quantum oscillator}

A complete orthonormal basis of $L_\vr^2$ can be obtained by solving the eigenvalue problem of the fractional harmonic oscillator. Set $\hbar=1$ and also $m=1=\om$. The following Hamiltonian is manifestly self-adjoint as sum of products of self-adjoint operators:
\be
\label{har}
\hat H= \tfrac12 \, \hat P^2 + \tfrac12\,  \hat Q^2\,.
\ee 
The eigenvalue problem 
\be
\label{har2}
-\frac{1}{2 \sqrt{v_i(x)} }\ \p_x^2\left[\sqrt{v_i(x)}\psi_n(x)\right] + \frac12 \,  x^2 \, \psi_n(x) = E_n \psi_n (x)
\ee
is trivially solved once recognized that Eq.\ \Eq{har2} is just the usual harmonic oscillator eigenvalue equation when written in terms of the function $\vp_n(x)=\sqrt{v_i(x)}\psi_n(x)$,
with normalized solutions and eigenvalues
\ba
\label{har3}
&& -\tfrac12 \ \p_x^2\varphi_n(x)  + \tfrac12 \, x^2 \, \varphi_n(x) = E_n \, \varphi_n(x)\,,\nonumber\\
&& \vp_n(x) = \frac{\rme^{-\frac{x^2}{2}}H_n(x)}{\sqrt{ \sqrt{\pi}2^n n!}}\,,\qquad E_n  = n+\frac12\,,
\ea 
where $H_n$ are Hermite polynomials 
\be
H_n(x):= (-1)^n \rme^{x^2} \p_x^n e^{-x^2}\,,\qquad n\in\mathbb{N}\,.
\ee
The eigenvalues $E_n$ and eigenfunctions $\psi_n$ can be also obtained via the method of ladder operators; thanks to the self-adjointness of $\hat{Q}$ and $\hat{P}$, in fact, it is possible to define consistently a set of creation, annihilation and number operators. The proof is identical to the standard case and we omit it. Since $\{\vp_n(x)\,|\,n\in\mathbb{N}\}$ is a complete orthonormal set in $L^2$, then 
\be
\label{har4}
\psi_n(x) =\frac{\varphi_n(x)}{\sqrt{v_i(x)}}= \frac{\rme^{-\frac{x^2}{2}}H_n(x)}{\sqrt{ v_i(x) \sqrt{\pi}2^n n!}}
\ee
is a complete orthonormal set in $L_\vr^2(x)$, and the $\psi_n$'s are the eigenfunctions of the Hamiltonian \Eq{har} with eigenvalues $E_n= n+1/2$. The Hamiltonian \Eq{har} is left invariant under the exchange $\hat P \leftrightarrow \hat Q$, so the corresponding equation in momentum space is identical, and its solutions are therefore the same functions \Eq{har4} with $x$ replaced by $k$: this means that the orthonormal set \Eq{har4} is left invariant by the  momentum transform $F_v$.

The wave-functions minimizing the uncertainty principle \Eq{heip} are weighted Gaussians:
\be\label{gau}
\psi(x)=\frac{\rme^{-\frac{(x-x_0)^2}{2}}}{\sqrt{v_i(x)\sqrt{\pi}}}\,,\qquad \|\psi\|^2=1\,.
\ee
Since $\langle\hat Q\rangle= x_0$, $\langle\hat Q^2\rangle=x_0^2+1/2$, $\langle\hat P\rangle= 0$, and $\langle\hat P^2\rangle= 1/2$, we obtain $(\Delta Q)^2(\Delta P)^2=1/4$. Note that for $x_0=0$ it gives just the lowest eigenfunction $\psi_0(x)$. In fractional spaces, the factor $[v_i(x)]^{-1/2}=[v_\a(x)]^{-1/2}\propto |x|^{(1-\a)/2}$ guarantees that the probability to find the particle at the origin, where the measure is integrably divergent, is zero.

It is important to stress that the shifted wave-function \Eq{gau} $\psi(x+x_*)$, where $x_*$ is some constant, no longer minimizes the uncertainty principle. This is not a surprise: translation invariance is broken in fractional theories. In spite of this fact, all the functions \Eq{gau} saturate the Heisenberg bound, for any value of $x_0$. This means that, although translation is no longer a symmetry, some similar symmetry survives. Indeed, it is the symmetry generated by the operator $\hat P$. Let us consider its action on the basis functions $\psi_n = \varphi_n(x)/\sqrt{v_i(x)}$:
\ba\label{har6}
\exp\bigl(-x_0 \hat P\bigr) \psi_n(x)& =& \sum_{m=0}^\infty\frac{(-x_0)^m \hat P^m}{m!} \ \frac{\varphi_n(x)}{\sqrt{v_i(x)}}\nonumber \\
 &=& \frac{1}{\sqrt{v_i(x)} } \sum_{m=0}^\infty \frac{(-x_0)^m}{m!} \p^m_x\left[ \sqrt{v_i(x)} \frac{\varphi_n(x)}{\sqrt{v_i(x)}}\right] \nonumber\\
 &=& \frac{1}{\sqrt{v_i(x)} }\  \exp\left[-x_0 \p_x\right] \varphi_n(x)\nonumber\\
 &=& \frac{\varphi_n(x-x_0)}{\sqrt{v_i(x)}}\,.
\ea
``$P$-translations'' work in a very convenient way: they first remove the translation non-invariant (and non-analytic) factor $[v_i(x)]^{-1/2}$, then act as a translation, and then put back the removed non-analytic factor. In other words, assuming that $v_i$ always depends on $|x|$, $P$-translations shift only the analytic part of the wave-functions. Thus, the system is not invariant under spatial translations but it is invariant under $P$-translations. The eigenfunctions of the fractional harmonic oscillator equation are $\psi_n(x;x_0) = \varphi_n(x-x_0)/\sqrt{v_i(x)}$, for any $x_0$, so that the Gaussians \Eq{gau} saturating the Heisenberg bound are nothing but the most general lowest-level eigenfunctions $\psi_0(x;x_0)$ (including $P$-translation degeneracy).

\begin{figure}
\centering
\includegraphics[width=7.8cm]{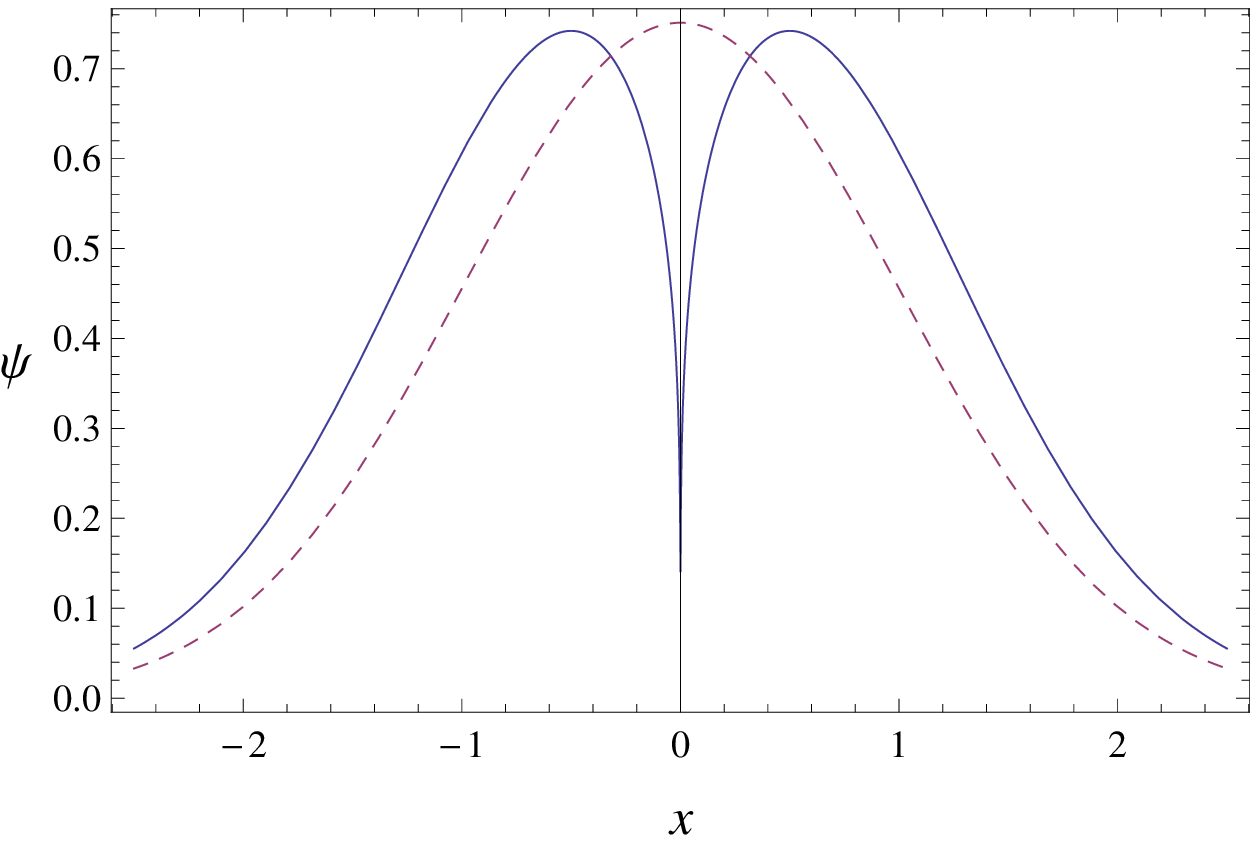} \hspace{.5cm}
\includegraphics[width=7.8cm]{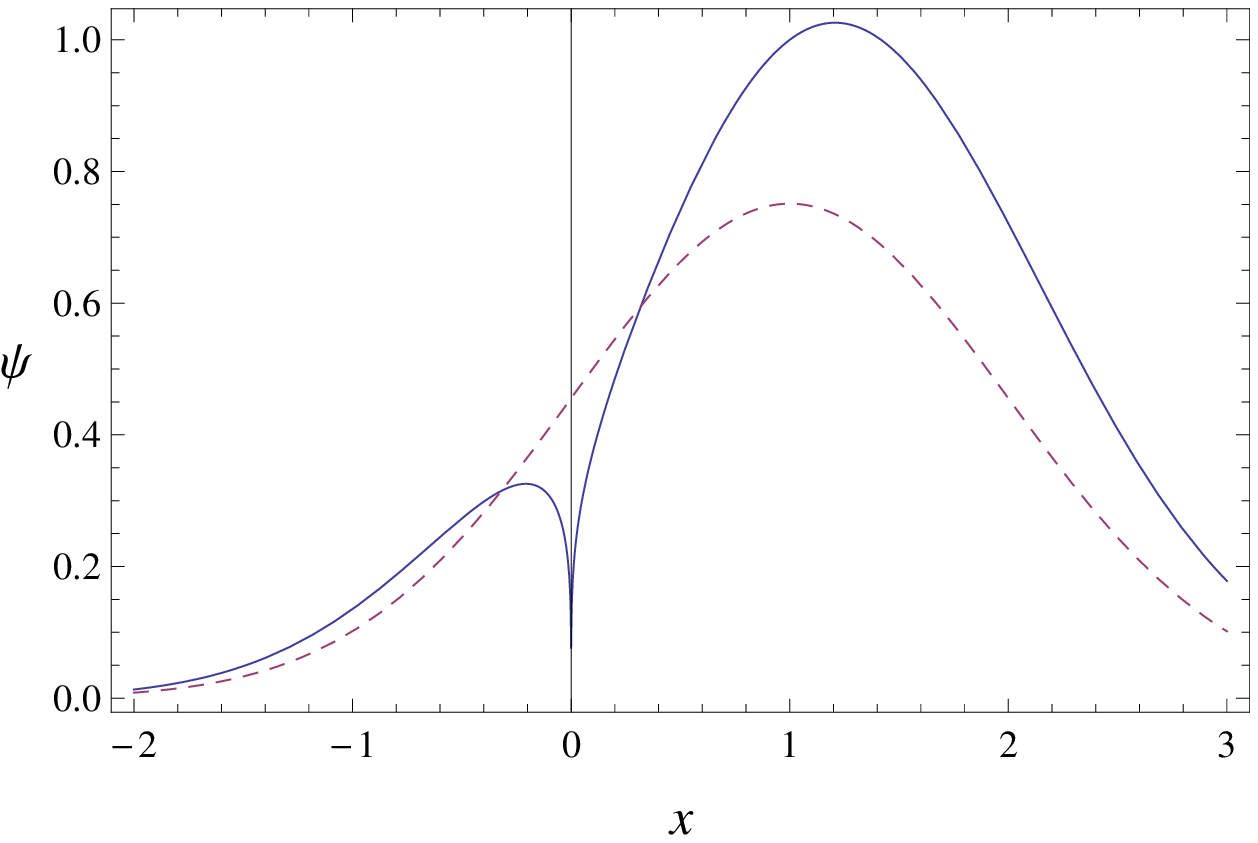}
\caption{\label{fig1}
The fractional wave-functions \Eq{gau} minimizing the uncertainty principle, peaked at $x_0=0$ (thick curve, left panel) and near $x_0=1$ (thick curve, right panel), for $\a=1/2$. The dashed curves are the Gaussian in the ordinary case, $\a=1$.}
\end{figure}

In the ordinary case, translational symmetry trivially corresponds to different choices of the origin of the coordinate system. In the fractional case, different choices of $x_0$ give different functions. In Figure \ref{fig1} the wave-function \Eq{gau} is depicted for $v_i=v_\a$ and the values $x_0=0$ and $x_0=1$; it vanishes at the origin, where it is not analytic (cusp). Contrary to the ordinary case, there are two local peaks, the one near $x_0$ being the higher. When $x_0=0$, both peaks have the same height and are symmetric with respect to the origin. These are two of the lowest eigenfunctions (parametrized by $x_0$) of the fractional harmonic oscillator, and both saturate the Heisenberg principle.

The presence of cusps is a clear evidence of non-analyticity of the wave-functions, which are not differentiable at the origin. However, ordinary derivatives are not the most natural differential operators in this fractional context, as they never correspond to hermitian operators. Only the combination defining $\hat P$ is hermitian and, regardless the presence of cusps, all the harmonic-oscillator wave-functions are $C^\infty$ with respect to the $P$-derivatives. 
This property is somewhat reminiscent of the fact that most fractal curves and other functions common in chaos theory are nowhere differentiable in the ordinary sense, but they can be differentiable in the sense of fractional calculus (e.g., \onlinecite{KoG1,KoG2}).

Once given the complete set of normalized eigenfunctions, we can compute the Green function using Eq.\ \Eq{gi}. To this purpose, one should first recall that, for any $|z|<1$ (Eq.\ 18.18.28 of Ref.\ \onlinecite{NIST}),
\be
\sum_{n=0}^\infty \psi_n (x)  \psi_n(y) z^n = \frac{1}{\sqrt{\pi \, v_i(x) v_i(y) (1-z^2)}}\, e^{[4 x y z-(x^2+y^2)(1+z^2)]/[2(1-z^2)]}.
\ee 
Then, applying the above formula with $z= e^{-i (t-t')/(\hbar + i \epsilon)}$ (in such a way that $|z|<1$) and eventually performing the limit $\epsilon \to 0$ we get
\be
G(x,t;x',t')= \sqrt{\frac{v_0(t')}{v_0(t) v_i(x) v_i (x')}} \, \,\sqrt{\frac{1}{2 \pi i \hbar \sin(t-t') }}\,  e^{\frac{i}{\hbar}  \frac{1 }{2 \sin (t-t')}  \left[(x^2 + x'^2) \cos (t - t') -2x x'\right]}.\label{hagr}
\ee
As in the free particle case (usual and fractional), the argument of the exponential is the classical action evaluated on the classical trajectory. Again, however, it should be stressed that the fractional action evaluated on the classical trajectory $q(t)$ is indistinguishable form the classical non-fractional action of the corresponding $\chi$-labelled system. This means that, as it is, the exponent in \Eq{hagr} is unable to discriminate between fractional and non-fractional systems. All the difference is in the term $\sqrt{v_0(t')/[v_0(t) v_i(x) v_i (x')]}$.


\section{Path integral}\label{path}

The path integral can be formulated in a fractional context. The Green function previously obtained shall be recovered as a path integral. This formalism will permit to clarify the role of the non-unitary pre-factor of the fractional theory, providing an interpretation of the latter as a topological term.


\subsection{Quantum mechanics from path integral}\label{pagr}

As is well known, the path integral picture of quantum mechanics relies on two postulates. It is worth repeating them to understand how these should be intended in a fractional world.
\begin{enumerate}
\item The Green function $G(x,t;x',t')$ of the system is a sum of phases,
\be
\label{pos1}
G(x,t;x',t')= C \sum_{\rm paths} e^{\frac{i}{\hbar} S[x,t;x't']}.
\ee
The righ-hand side of this equation is entirely classical. It is a sum over all possible classical trajectories joining the point $x'$ at the time $t'$ with the point $x$ at the time $t$. The argument of the exponential is the classical action evaluated at that particular trajectory. All the trajectories have the same weight.
\item
$C$ is a constant to be determined in such a way that $G$ behaves as a probability density: for any intermediate $\bar t\in[t',t]$, one must have
\be
\label{pos2}
G(x,t;x',t')=  \sum_{\bar x} G(x,t;\bar x,\bar t) G(\bar x,\bar t;x',t')\,.
\ee
\end{enumerate}
Let us begin from \Eq{pos2}. If the space has anomalous measure $\vr(x)$, the sum over the intermediate positions $\bar x$ should be interpreted as an integral with measure $d\vr(\bar x) = d\bar x\, v_i(\bar x)$. But then, the identity Green function (corresponding to no propagation) should be realized by a fractional delta, Eq.\ \Eq{deltafrac}. In turn, the identity Green function is attained for $t=t'$ and then 
$\bar t=t'$ in Eq. \Eq{pos2}. Consequently, this equation can be interpreted as a resolution of the identity when evaluated at $t=\bar t = t'$. 
Comparing this with \Eq{fbmt}, whose integral over $dp$ provides a resolution of the identity, one can conclude that all the $(x,x')$ dependence in $C$ is of the form
\be
C= \frac{C(t',t)}{\sqrt{v_i(x) v_i (x')}}\,, \quad \quad C(t,t)= {\rm const.}
\ee
 where $C(t',t)$ is a function  that can depend on time both explicitly and implicitly, through $v_0(t)$. The latter dependence can be easily recovered by simple scaling arguments. From now on we shall specialize to the case of quadratic Lagrangians, so that all the $v_0(t)$ factors can be reabsorbed with the redefinition $\chi(t) = \sqrt{v_0(t)} \, x(t)$, while the whole $x$ dependence in the action is through $\chi$ and its time derivatives. For instance, in the free-particle case
\be
\frac12 m \int_{t'}^t \rmd \tau  \, v_0(\tau) \, [\cD_\tau x(\tau)]^2 = \frac12 m \int_{t'}^t \rmd \tau\  [\dot \chi (\tau)]^2\,.
\ee
Consequently, in the small $t-t'$ limit the resolution of the identity can only provide a $\delta (\chi)$, which scales as $\delta(x)/\sqrt{v_0(t)}$. Hence the implicit $t$ dependence of
$C(t',t)$ is $1/\sqrt{v_0(t)}$. In turn, this fixes the $v_0(t')$ dependence as $\sqrt{v_0(t')}$, in such a way the $v_0$ dependence cancels at $t=t'$.
Summarizing, initial conditions and scaling arguments fix the following form of $C$:
\be
\label{const1}
C= { K}\  \sqrt{\frac{v_0(t')}{v_0(t) v_i(x) v_i (x')}}\,, 
\ee
where $K$ is constant both in space and time.

Although most of the considerations will hold for any quadratic Lagrangian, we shall continue for simplicity with the free-particle case. The sum over all the possible paths in \Eq{pos1} can be conveniently performed by slicing the time interval  $t-t'$ 
into $N$ infinitesimal  parts,
\ba  
&&t=t_N>t_{N-1}>t_{N-2}>\dots>t_2>t_1>t_0=t'\,,\nonumber\\
&& t_n-t_{n-1}=\epsilon\,,\quad N\epsilon=t-t'\,,\label{split}
\ea
and then enacting the standard discretization of variables $x_n=x(t_n)$. Notice that this is the ordinary splitting except for the requirement that none of the points coincide with the singularity of the measure; in the fractional case, $t_n\neq 0$. While this constraint on the internal points is immaterial due to the eventual analytic continuation leading to the final result, it still holds for the initial and final points $t$ and $t'$. The singularities in the measure must not correspond to either boundary point. Apart from this, there is no other complication and we do not even need to redefine time as in other path integrals featuring integral singularities, like the hydrogen-atom system.\cite{DuK} We shall evaluate first the discretized Green function $G_N(x,t;x',t')$, performing eventually the $N\to \infty $ limit. The  constant $K$ in \Eq{const1} is also defined as a limit, $K = \lim_{N\to \infty} \kappa_N$, $\kappa_N$ being the corresponding constant in the discretized Green function which, by virtue of \Eq{pos2}, can only  be a pure power,  $\kappa_N = (\kappa)^N$.

 Within the $n$-th slice, the action attains the value 
 \be
\frac{ i S_n}{\hbar} = \frac{im}{2\hbar \epsilon}\Bigl[\sqrt{v_0(t_n)} x_n-\sqrt{v_0(t_{n-1}) }x_{n-1}\Bigr]^2.
\ee
Iterating $N-1$ times Eq. \Eq{pos2}, we obtain
\begin{widetext}
\ba
\label{gdisc1}
G_N(x,t;x',t')&=& \kappa^{N}\sqrt{\frac{v_0(t_0)}{v_0(t_N)}}\int\frac{d\vr(x_1)\ldots d\vr(x_{N-1}) \ e^{\frac{i}\hbar\sum_n S_n}}{\sqrt{v_i(x_N)v_i(x_{N-1})}
\cdots \sqrt{v_i(x_1)v_i(x_{0})}}\nonumber\\
&=& \kappa^N \sqrt{\frac{v_0(t_0)}{v_0(t_N) {v_i(x_N)v_i(x_{0})}} }  \ \int dx_1\cdots dx_{N-1}  e^{\frac{i}\hbar\sum_n S_n} \nonumber\\
&=&\frac{\kappa^{N}}{{\sqrt{{\cal V}_N}}}\sqrt{\frac{v_0(t_0)}{v_0(t_N) v_i(x_N)v_i(x_{0})}}\int d\chi_1\cdots d\chi_{N-1}\,\exp\left[ \frac{im}{2\hbar \epsilon} \sum_{n=1}^{N} (\chi_n-\chi_{n-1})^2\right],
\ea
\end{widetext}
where we set $\chi_0=\sqrt{v_0(t_0)}\, x_0 \equiv \chi'$ and $\chi_N=\sqrt{v_0(t_N)}\, x_N \equiv \chi$.
Note that all the $v_i$ factors cancelled pairwise, except the first and the last.
The integration measure simplifies, and all the integrals are Gaussian. Positivity of the measure is instrumental to perform all simplifications without picking up phases. In the last equality, we performed the change of variables $\chi_n = \sqrt{v_0(t_n)}\, x_n$, with Jacobian 
\be
 \frac{1}{{\sqrt{{\cal V}_N}}}:=
\prod_{n=1}^{N-1} \frac{1}{\sqrt{v_0(t_n)}}\,.
\ee
Here ${\cal V}_N$ is the regularized (discretized) determinant of the operator defined by the kernel ${\cal V}(t,\tau)= v_0(t) \delta(t-\tau) $ in the interval $(t_0,t_N)$. Its $N\to \infty$ limit is either 1 or an ill-defined constant (zero or infinity) that will  be reabsorbed in the  constant $\kappa^N$ prior to the $N\to \infty$ limit. The remaining Gaussian integrals are straightforward, leading to 
\ba
G_N(x,t;x',t') &=& G_N(x_N,t_N;x_0,t_0)\nonumber\\
&=& \left(\frac{\kappa^2 2 \pi \epsilon \hbar}{i m {\cal V}_N^{1/N}}\right)^{\frac{N}{2}}\sqrt{\frac{im}{2\pi\hbar N \epsilon}} \sqrt{\frac{v_0(t_0)}{v_0(t_N) v_i(x_N)v_i(x_{0})}}\ 
e^{\frac{im(\chi_N-\chi_{0})^2}{2\hbar N\epsilon}}.\nonumber\\\label{gdisc2}
\ea

In taking the limit $N\to \infty$ in Eq.\ \Eq{gdisc2}, one should recall that $N\epsilon = t-t'$, so that the argument of the exponent
is just the action of the classical non-fractional particle evaluated on the classical solution (a particle with constant velocity $w=(\chi-\chi')/(t-t')$).
The only delicate term is the first one. Its large $N$ limit is undefined, unless 
\be
\label{kappaval}
\frac{\kappa^N}{{\sqrt{{\cal V}_N}}}=\left( \frac{im}{2\pi\hbar \epsilon} \right)^{N/2}.
\ee
 For such a value, 
the first factor gives $1$ for any $N$, and the Green function of the free particle exactly coincides with Eq.\ \Eq{gfree1},
\ba 
G(x,t;x',t')&:=&\lim_{N\to\infty}G_N(x,t;x',t')\nonumber\\
&=& \sqrt{\frac{v_0(t')}{v_0(t) v_i(x)v_i(x')}}\ \sqrt{\frac{im}{2\pi\hbar (t-t')}}\ e^{\frac{im(\chi-\chi')^2}{2\hbar(t-t')}},\label{gdisc3}
\ea
where, we recall, $\chi=\sqrt{v_0(t)}\,x$ and $\chi'=\sqrt{v_0(t')}\,x'$. In similar fashion, it can be shown that also the Green function \Eq{hagr} of the fractional harmonic oscillator can be obtained along the same lines.

\subsection{Non-unitarity and topological considerations}\label{topo}

Whatever the method chosen to evaluate the Green function (either via Eq.\ \Eq{gi} or through the path integral), we encountered the following general property: all the non-unitarity of the theory is encoded in the pre-factor $\sqrt{v_0(t')/v_0(t)}$ appearing in Eqs.\ \Eq{gfree1} (or \Eq{gdisc3}) and \Eq{hagr} and in the general expression \Eq{gi} (the $v_i$ terms, instead, provide the correct  normalization factors for the $L^2_\vr$ space). Were it not for such a term, the theory would be unitary and indistinguishable from the corresponding ordinary one. As we are going to see now, understanding the origin of this pre-factor also clarifies the main difference between ordinary and fractional theory at the quantum level, and the reason why classical free theories are equivalent.

The non-unitary term can be rewritten as an additional exponential of a time integral ($b\neq 0$ is an arbitrary constant)
\ba
\sqrt{\frac{v_0(t')}{v_0(t)}} &=& e^{-\{\ln [b \sqrt{v_0(t)}]- \ln[ b \sqrt{v_0(t')}]\}}\nonumber\\
&=& e^{-\int_{t'}^t d\tau \p_\tau \ln [b \sqrt{v_0(\tau)}]}\nonumber\\
&=& e^{- \frac1b \int_{t'}^t d\tau \cD_\tau(b)}\,.\label{tode}
\ea
In the non-fractional case, the Green function of a free particle is the exponential of $(i/\hbar) \bar S$, $\bar S$ being the classical action evaluated on the classical solution. In the fractional theory, the classical action $S_{\rm cl}$ evaluated on the classical solution is identical to the free case, $S_{\rm cl}=\bar S$ (see Eqs.\ \Eq{gfree1} and \Eq{hagr}), so the same classical action cannot serve as exponent of the quantum  Green function also in the fractional context. The crucial and \emph{only} discriminator between fractional and ordinary free theory is the total derivative \Eq{tode}. In fact, the non-unitary pre-factor needed to obtain the correct Green function can be completely reabsorbed in a redefinition of the action:
\be\label{cocy}
S_{\rm cl}\longrightarrow S_{\rm qu}= S_{\rm cl} +\frac{i\hbar}{b}\int_{t'}^t d\tau\, \cD_\tau (b)\,.
\ee
The difference between $S_{\rm cl}$ and $S_{\rm qu}$ is a boundary term and the two actions are classically equivalent (their difference is an integral of a total derivative that does not contribute to the equations of motion). Quantum mechanically, they are \emph{not}: both of them describe free theories, but the former in a standard spacetime, the latter in a fractional one. Therefore, all the effect of a fractional spacetime can be encoded in a purely topological term.

To summarize, the traditional recipe $G\sim e^{(i/\hbar)\bar S}$ for the path integral is incomplete inasmuch as two systems having the same classical action on classical trajectories may be still physically inequivalent quantum mechanically, and they can be distinguished by a topological term. This term strongly reminds us of a 1-cocycle,\cite{Jac84} although its interpretation is different. While 1-cocycles arise whenever the action possesses a symmetry not enjoyed by the Lagrangian and they enter as projective representations of such symmetry, in this case the topological contribution tells apart ordinary from fractional quantum systems.



\subsection{Path integral from quantum mechanics}

The purpose of this section is complementary to the previous ones: we wish to recover the path integral postulate \Eq{pos1} entirely from a quantum mechanical setting. We fix $m=1$.

Let us follow the standard way by considering a time interval and subdividing it into $N$ small parts as in Eq.\ \Eq{split}.
Recalling Eq.\ \Eq{xUx}, we introduce $N-1$ completeness relations
\be\label{core}
\int \rmd x\,v_i(x) \ket{x}\bra{x}=\mathbbm{1}\,.
\ee
Then, the time evolution operator must be factorized as
\be\label{Usp}
\hat{U}(t,t')= \frac{1}{\sqrt{v_0(t)}}e^{-\frac{i}{\hbar}\hat{H}\epsilon}e^{-\frac{i}{\hbar}\hat{H}\epsilon}\dots\,e^{-\frac{i}{\hbar}\hat{H}\epsilon}\sqrt{v_0(t')}\,,
\ee
and it is clear that the following formula holds:
\be 
G_N(x,t;x',t')= \sqrt{\frac{v_0(t')}{v_0(t)}}\int \prod_{s=1}^{N-1} \rmd x_s\,v_i(x_s)\prod_{n=0}^{N-1} G_{n+1,n}(x_{n+1},x_n,t_{n+1},t_n)\,,\nonumber\\\label{green}
\ee
where
\ba
G_{n+1, n}&=&\left\langle x_{n+1}\left|e^{-\frac{i}{\hbar}\hat{H}\epsilon}\right|x_n\right\rangle\nonumber\\
&=&\left\langle x_{n+1}\left|\left(\mathbbm{1}-i\frac{\hat{H}\epsilon}{\hbar}+\dots\right)\right|x_n\right\rangle\,.
\ea
Consider a Hamiltonian operator of the form $\hat{H}= \hat{P}^2/2+V(\hat{Q})$. Then,
\ba
\braket{x_{n+1}|\hat{H}|x_n}&=&\left\langle x_{n+1}\left|\tfrac{1}{2}\hat{P}^2+V(\hat{Q})\right|x_n\right\rangle\nonumber\\
&=&\int \rmd p\,v_i(p)\braket{x_{n+1}|p}\left\langle p\left|\tfrac{1}{2}\hat{P}^2+V(\hat{Q})\right|x_n\right\rangle\nonumber\\
&=&\int \frac{\rmd p}{2\pi\hbar} \frac{e^{\frac{i}{\hbar}p(x_{n+1}-x_n)}}{\sqrt{v_i(x_{n+1})v_i(x_n)}} \left[\frac{1}{2}p^2+V(x_n)\right],\nonumber\\
\ea
where we used Eq.\ \Eq{aval1} for the eigenfunctions of the fractional momentum operator $\hat{P}$. Thus, we have
\ba
G_{n+1,n}&=&\int \frac{\rmd p}{2\pi\hbar} \frac{e^{\frac{i}{\hbar}p(x_{n+1}-x_n)}}{\sqrt{v_i(x_{n+1})v_i(x_n)}} \left[1-i\frac{H(p,x_n)\epsilon}{\hbar}+\dots\right]\nonumber\\
&=&\int \frac{\rmd p}{2\pi\hbar}\frac{e^{\frac{i }{ \hbar}\left[ p(x_{n+1}-x_n) - \epsilon H(p,x_n)\right]}}{\sqrt{v_i(x_{n+1})v_i(x_n)}}\,.
\ea

The various spatial measures in Eq.~\Eq{green} simplify  except for  those corresponding to $(x,t)$ and $(x',t')$, and we have
\begin{widetext}
\ba
G_N(x,t;x',t') &=& \sqrt{\frac{v_0(t')}{v_0(t) v_i(x) v_i(x')}}\ \frac{1}{(2\pi \hbar)^N}\ \int dx_1\cdots dx_{N-1} dp_1 \cdots dp_N\nonumber \\
&&\times \exp\left\{\frac{i \epsilon}{\hbar}\sum_{n=0}^{N-1} \left[\frac{p_{n+1}(x_{n+1}-x_n)}{\epsilon}- \frac{p_{n+1}^2}{2 m} - V(x_n) \right]\right\}.\label{bah}
\ea
Integrations over momenta are Gaussian, giving
 \ba
 G_N(x,t;x',t')&=& \sqrt{\frac{v_0(t')}{v_0(t) v_i(x) v_i(x')}}\ \left( \frac{ m i}{2\pi \hbar \epsilon}\right)^{N/2}\ \int dx_1\cdots dx_{N-1}\exp\left\{\frac{i \epsilon}{\hbar}\sum_{n=0}^{N-1} \left[\frac{m(x_{n+1}-x_n)^2}{2\epsilon^2} - V(x_n) \right]\right\}\nonumber \\
&=& \sqrt{\frac{v_0(t')}{v_0(t) v_i(x) v_i(x')}}\ \kappa^N \ \int dq_1\cdots dq_{N-1}\nonumber \\
&&\times \exp\left(\frac{i \epsilon}{\hbar}\sum_{n=0}^{N-1} \left\{\frac{m[\sqrt{v_0(t_{n+1})}\, q_{n+1}-\sqrt{v_0(t_n)}\, q_n]^2}{2\epsilon^2} - v_0(t_n) V(q_n) \right\}\right),
\ea
where, in the last equation, we used Eq.\ \Eq{kappaval} and changed variables $x_n = \sqrt{v_0(t_n)}\, q_n$. The Jacobian of the change of variables is precisely that in \Eq{kappaval}, and we used the fact that the potential is quadratic to scale a $v_0$ factor out of $V$.

We now perform the large $N$ limit to get the Green function $G(x,t;x',t')$. In this limit,  the first factor gives \Eq{const1}, the differentials $\prod_n dq_n$ provide the definition of the functional measure $[dq]$, and the action in the exponential is just the fractional action, 
\be
G(x,t;x',t')= K \ \sqrt{\frac{v_0(t')}{v_0(t) v_i(x) v_i(x')}}\int[dq]\, \exp\left\{\frac{i}{\hbar}\int_{t'}^t dt \, v_0(t) \left[\frac12 m (\cD_t q)^2-V(q)\right]\right\},
\ee
\end{widetext}
in full agreement with Sec.\ \ref{pagr}. For non-quadratic Lagrangians the problem is subtler but, in those cases, the path integral is typically unsolvable.

Notice that the last line of Eq.\ \Eq{bah} defines the relation between Hamiltonian and Lagrangian. The discrete expression $\e \bar L_n=p_{n+1}(x_{n+1}-x_n)- \e H_n$, $H_n:=p_{n+1}^2/(2m)-V(x_n)$, translates into the Legendre transform $\bar L=p \dot x-H$, where $\bar L$ is the ordinary Lagrangian. Changing variables before integrating over the momenta, this relation becomes $L=p \cD_tq-H$ for the fractional theory, in agreement with the \emph{Ansatz} \Eq{H-clas}.


\section{Discussion}\label{conc}

Spaces with non-trivial but factorizable measures provide an interesting arena where to explore avenues towards ultraviolet-complete models of Nature. All questions asked in standard scenarios of space, time and matter should be addressed also in these anomalous-spacetime models. As an application of the existence of unitary invertible transforms \cite{frc3} between configuration and momentum fractional spaces, we have constructed a well-defined quantum mechanics on such spaces, proving Heisenberg's principle, formulating the path integral and considering the standard examples of the free particle and the harmonic oscillator. In the latter case, the energy levels of the Hamiltonian are the usual ones but the eigenfunctions of the problem are augmented by a factor dependent on the Hausdorff dimension of the space. These wave-functions have several unusual properties. They are ``$P^\infty$,'' i.e., non-analytic in the origin but everywhere differentiable, infinitely many times, with respect to the first-order fractional momentum operator $\hat P$. This operator generates a modification of ordinary translations which do not leave invariant the shape of the wave-functions, yet it transforms one into another with the same Heisenberg uncertainty. Similarly, the time evolution operator realizes a fractional modification of ordinary time translations. In general, the theory is non-unitary due, however, to a topological effect dependent on the initial and final state. This effect, which tells apart fractional from standard quantum dynamics, clearly arises from a topological term in the path-integral formulation.

Before concluding, we examine an issue related to the fractal character of quantum mechanical paths. As we recalled in the introduction, the path of a free quantum particle is continuous but nowhere differentiable.\cite{FeH} This property, together with their self-affinity (in general, these curves are not self-similar), counts quantum paths as random-fractal curves. At scales larger than the Compton wavelength of the particle but smaller than the de Broglie wavelength, in $D=1$ the graph (i.e., the plot of ${\bf q}(t)$ versus $t$) of a quantum-mechanical non-relativistic particle has Hausdorff dimension $\dh=3/2$, while the path (i.e., the ``trace'' left by the particle in configuration space ${\bf q}$) has $\dh=1$. In $D\geq 2$, both have $\dh=2$.\cite{AbW,CROS,AHM,Dew90,Sor90,KLMP,Kro97,NiN} At smaller scales $\dh=1$, as one can understand from the analogy between quantum paths and Brownian motion.\cite{AHM,Sor90,Nel66} The relativistic particle was studied as well, and $\dh=1$.\cite{CaF} Given the intrinsic ``fractal'' nature of the classical spacetime wherein the particle moves in fractional theories, one could wonder how the above results on the Hausdorff dimension of graphs and paths are altered. The answer is that they are not altered at all. One can show this by an explicit calculation following, e.g., the one in Ref.\ \onlinecite{AbW} based on wave-functions and the time evolution operator. One starts from the operational definition of fractal dimension (which, in most cases, coincides with the Hausdorff dimension $\dh$). Since fractional spacetimes are continuous, this ``fractal dimension'' is indeed $\dh$ and we can simply refer to it as the Hausdorff dimension. Let $l_\Delta$ be the ordinary length of a curve measured with a probe with some given finite resolution $\Delta x\neq 0$. The Hausdorff length $\lh$ of the curve is defined as $\lh=\lim_{\Delta x\to 0} l_\Delta (\Delta x)^{\dh-1}$. For a smooth curve, $\dh=1$ and the Hausdorff and ordinary lengths coincide, while for fractal curves only $\lh$ is a meaningful geometric indicator. In fact, as the resolution $\Delta x$ decreases $l_\Delta$ tends to infinity (because one needs, say, more and more balls to cover the curve); the Hausdorff dimension is always larger than the topological dimension of the fractal set (in this case, 1). (On the other hand, $\dh$ cannot be greater than the topological dimension of the space wherein the curve is embedded.) Requiring that $\lh$ be resolution independent yields the correct value of the Hausdorff dimension; the example of the von Koch curve can be found, e.g., in Ref.\ \onlinecite{NiN}.

For a quantum particle, $l_\Delta$ is replaced by the average distance $\langle l_\Delta\rangle$ covered by the particle in a time interval $\Delta t=t-t'$. The average is calculated on a state  associated with a sequence of position measurements localizing the particle in a region of size $\Delta x$. The only difference, immaterial for the final result, with respect to the standard case is in the non-unitary pre-factor $v_0(t')/v_0(t)$ in the expression for $\langle l_\Delta\rangle$. Up to this factor, clearly inherited from the fractional operator $\hat U$ (appearing twice in the expectation value), the scaling of the average distance is the standard one, $\langle l_\Delta\rangle\propto \Delta t/\Delta x$. Replacing this into $\lh$ and demanding the Hausdorff length to be resolution-independent yields $\dh=2$. This result was expected because the quantum dynamics of the free particle is the same in ordinary and fractional spacetimes except for a boundary term. The latter, of course, does not modify the local structure of quantum paths, hence the conclusion.

A viable construction of quantum mechanics on fractional spacetimes opens up the possibility to test the latter as models of Nature. As stated in the Introduction, eventually one would like to make predictions in a field-theory context where the dimension of spacetime changes with the scale. However, at large-enough space/time scales the effect of dimensional flow is small and one can consider a fractional spacetime with fixed dimensionality close to $3+1$, corresponding to an expansion in the small parameters $\e_{0,i}\sim 1-\a_{0,i}$. The scales at which this approximation holds are at least atomic,\cite{frc2} possibly smaller. It is in this context that the framework presented here can help to place bounds on the theory from experiments in quantum mechanics. For example, the difference between the theoretical predictions and experimental data for the Lamb shift in hydrogen can be ascribed to dimensional effects in order to give an upper bound for $\e$. For theories with naive dimensional-regularization-like modifications, the constraint is $|\e|<10^{-11}$ at scales $\ell\sim 10^{-11}\,{\rm m}$.\cite{ScM,MuS} It also turns out that this upper bound is more stringent than others at electroweak or astrophysical scales.\cite{frc2} A revisit of the same problem in fractional spacetimes should be able to yield a similar constraint, probably of the same order of magnitude. This will be the subject of future investigations.


\begin{acknowledgments}
{\small G.C.\ and M.S.\ acknowledge financial support through a Sofia Kovalevskaja Award at the AEI.}
\end{acknowledgments}


\end{document}